\documentclass[sigconf]{acmart}

\usepackage{booktabs} % For formal tables
\usepackage{enumitem}

\setcopyright{rightsretained}
\usepackage{tabu}

\acmDOI{10.475/123_4}

\acmISBN{123-4567-24-567/08/06}

\copyrightyear{2017} 
\acmYear{2017} 
\setcopyright{acmcopyright}
\acmConference{DLRS 2017}{August 27, 2017}{Como,
Italy}\acmPrice{15.00}\acmDOI{10.1145/3125486.3125489}
\acmISBN{978-1-4503-5353-3/17/08}

\begin{document}
\title{Specializing Joint Representations for the task of Product Recommendation}
%\subtitle{Extended Abstract}
%\subtitlenote{The full version of the author's guide is available as
%  \texttt{acmart.pdf} document}

\author{Thomas Nedelec}
\affiliation{\institution{Criteo Research}}
\email{t.nedelec@criteo.com}

\author{Elena Smirnova}
\affiliation{\institution{Criteo Research}}
\email{e.smirnova@criteo.com}

\author{Flavian Vasile}
\affiliation{\institution{Criteo Research}}
\email{f.vasile@criteo.com}


\begin{abstract}
We propose a unified product embedded representation that is optimized for the task of retrieval-based product recommendation. To this end, we introduce a new way to fuse modality-specific product embeddings into a joint product embedding, in order to leverage both product content information, such as textual descriptions and images, and product collaborative filtering signal. By introducing the fusion step at the very end of our architecture, we are able to train each modality separately, allowing us to keep a modular architecture that is preferable in real-world recommendation deployments. We analyze our performance on normal and hard recommendation setups such as cold-start and cross-category recommendations and achieve good performance on a large product shopping dataset. 
\end{abstract}

 \begin{CCSXML}
<ccs2012>
<concept>
<concept_id>10010147.10010257</concept_id>
<concept_desc>Computing methodologies~Machine learning</concept_desc>
<concept_significance>300</concept_significance>
</concept>
<concept>
<concept_id>10010147.10010257.10010293.10010294</concept_id>
<concept_desc>Computing methodologies~Neural networks</concept_desc>
<concept_significance>300</concept_significance>
</concept>
</ccs2012>
\end{CCSXML}

\ccsdesc[300]{Computing methodologies~Machine learning}
\ccsdesc[300]{Computing methodologies~Neural networks}

% We no longer use \terms command

\keywords{Recommender systems, representation learning, embeddings, second-order interactions}

\maketitle

\section{Introduction}
\label{sec:intro}
Online product recommendation is now a key driver of demand, not only in E-commerce businesses that recommend physical products, such as Amazon \citep{amzreco}, TaoBao \citep{taobaoreco} and Ebay \citep{ebayreco}, but also in online websites that recommend digital content such as news (Yahoo! - \cite{agarwal2013content}, Google - \cite{liu2010personalized}), movies (Netflix - \cite{bell2007lessons}), music (Spotify - \cite{johnson2015algorithmic}), videos (YouTube - \cite{covington2016deep}) and games (Xbox -  \cite{koenigstein2012xbox}).

One of the most popular architectures for recommendation at scale (see \cite{criteoreco}, \cite{cheng2016wide}, \cite{covington2016deep}) divides the recommendation process in two stages: \emph{a candidate generation stage} that prunes the number of recommendable items from a volume of potentially billions of items to a couple of hundreds, followed by a second \emph{item selection stage} that decides the final set of items to be displayed to the user,(see Figure \ref{fig:criteo_recsys} in the Appendix for more details).

Due to lower constraints on the latency and the potential impact resulting from an improvement on the candidate generation stage, we choose to concentrate our efforts on the task of optimal candidate generation. We formalize the problem as a link prediction task, where given a set of past pairs of co-purchased products we try to predict the probability of being cobought for unseen pairs of products. Related work in representation learning for recommendation investigated the use of both collaborative filtering \citep{prod2vec} and content information \citep{mcauley2015image}, but to our knowledge, there has been no attempt to unify them in a single representation. We see this as an opportunity to investigate the leveraging effect of generating a \emph{Specialized Joint Representation} via a deep-learning approach. 

In order to achieve this, we propose Content2Vec -  a modular deep architecture that leverages state-of-the-art architectures for generating embedded representations for image, text and collaborative filtering (CF) input, re-specializes the resulting product embeddings and combines them into a single product vector. This is a very general architecture that can plugin any neural networks for modeling the input information and re-use them for the problem of product recommendation. 

%Modular architecture
We argue that a modular architecture coupled with a module-by-module training is the easiest way to put such a complex model in production. In Content2Vec, most of the computation is spent on modeling the modality-specific representations, which in the case of periodic retraining should be able to leverage the previous models as the initialization states and converge very fast on the new input data. Furthermore, using pre-trained models for each modality allows us to leverage external sources of data and do transfer learning, whose value was repeatedly confirmed(\cite{yosinski2014transferable},\cite{veit2015learning}). 
\\
\\
In the following, we formally define the set of associated requirements that define an optimal product embedding:   
\begin{itemize}[topsep=6pt]
\item \textbf{Relevance}: the representation should be optimized for product recommendation relevance, as measured by the associated target metrics (modeling it as a link prediction task and optimizing for the AUC of product pair prediction).
\item \textbf{Coverage}: the representation should leverage all available product information (in our case, all product information available in the product catalog together with observed product co-occurrences).
\item \textbf{Cross-modality expressiveness}: the representation should be able to account for interactions between various information sources such as text and image (can take into account the fact that the word "red" and the "red" color detector are correlated).   
\item \textbf{Robustness}: the representation should operate well (recommendation performance will not degrade dramatically) in hard recommendation situations such as product cold-start (new products, new product pairs) and cross-category recommendation. These are important use-cases in product recommendation, when the product catalog has high churn (as in the case of flash sales websites or classifieds) or the recommendation needs to leverage cross-advertiser signal (as in the case of new users and user acquisition advertising campaigns). This is a different goal from simply trying to optimize for relevance metrics, due to the inherent limitations of offline metrics in predicting future online performance.
\item \textbf{Retrieval-optimized}: the representation should be adapted to a content-retrieval setup, both on the query and on the indexing side, meaning that the vectors should be either small, sparse or both.
\end{itemize}
%Experiments: 
We analyze the performance of our proposed architecture on the five requirements presented above on an Amazon dataset \citep{mcauley2015image} that contain information on co-purchased products. We report our improvements versus text, image (\citep{mcauley2015image}) and collaborative-filtering (\citep{prod2vec}) based baselines - and a simple combination of the three. We show improvements both on normal and hard recommendation regimes such as cold-start and cross-category setups. 

%Contributions: 
Our main contributions are the following:
\begin{itemize}[topsep=6pt]
\item We propose a novel way of integrating deep-learning item representation in the context of large scale recommender system with a two-stage serving architecture and introduce the new task of \emph{Specialized Joint Representation} for optimal candidate selection in both cold start and normal recommendation setups.
\item We introduce a new deep architecture that merges content and collaborative filtering signal for the task of product recommendation and propose the \emph{Cross Interaction Unit} (CIU), a new learning component that models the joint product representation. We benchmark the different architectures in two experimental setups (hard cold start, cross-category) that test the robustness of our architecture. 
\end{itemize}
%\item We introduce an explicit way to generate cross-features across product modalities using CNNs
%\item We develop a solution that approximates product-product non-linearities with product non-linearities. to be able to use the models directly with a KNN-based candidate retrieval model and investigates sparse variants.
%Though the focus of our work is on improving product recommendation through representation learning, we believe that simple extensions of our approach can be applied to many other recommendation scenarios by incorporating additional input neural network modules specific to the available signal (such as video or audio signal).
The rest of the paper goes as follows: In Section \ref{sec:relatedwork}, we cover previous related work and the relationship with our method. In Section \ref{sec:model}, we present an overview of our new architecture and how we learn to compute similarities between products. 

Section \ref{sec:Content-specific embedding modules} details the Content-specific embedding modules that are used to build several representations from the different modalities: text, image and collaborative filtering data. In Section \ref{sec:architecture:product}, we propose several architectures to fuse the modality-specific signals and build a unified product embedding. In Section \ref{sec:experimentsetup}, we present the experimental setup and go over the results in Section \ref{sec:results}. In Section \ref{sec:conclusions}, we summarize our findings and conclude with future directions of research.
\section{Related Work}
\label{sec:relatedwork}
Our work fits in the new wave of deep learning based recommendation solutions, that similarly to classical approaches can fall into tree categories, namely collaborative filtering based, content based or hybrid approaches.
\subsection{Collaborative filtering methods}
Several approaches use neural networks to build better item representations based on the co-occurrence matrix. The Prod2Vec algorithm \citep{prod2vec} implements Word2Vec (\citep{Word2Vec}, \citep{shazeer2016swivel}), an algorithm that is at origin a shallow neural language model, on sequences of product ids, to reach a low-dimensional representation of each product. Among other embedding solutions that use the item relationship graph are the more recent extensions to Word2Vec algorithm such as Glove \citep{glove}, SWIVEL \citep{shazeer2016swivel}, the graph embedding solutions proposed in Node2Vec \citep{grover2016node2vec} and SDNE \citep{wang2016structural}.
\subsection{Content based methods}
Content-based methods recommend an item to a user based upon an item description and a user profile (\citep{pazzani2007content}). This idea was deeply investigated in the information retrieval literature: in the context of web search, DSSM \citep{huang2013learning} and its extensions \citep{shen2014learning}(C-DSSM) and \citep{shan2016deep} are some of the successful methods that specialize query and document text embedding in order to predict implicit feedback signal such as document click-through rate. In the context of product recommendation, \citep{mcauley2015image} feed a pre-trained CNN with products images. The network was pretrained on the ImageNet dataset, an image classification task that is very different from the task of image-based product recommendation. The last layer of the network is used as the product embedding. This representation is subsequently used to compute similarities between products. Similarly, the authors in \citep{van2013deep} use CNNs to compute similarities between songs. \cite{yosinski2014transferable} the authors show that the low layers of DNNs trained on different tasks are often similar and that good performance can be reached by fine-tuning a network previously trained on another task. In the case of recommendation systems, this fine tuning was implemented in \cite{veit2015learning}, where the authors specialize a GoogLeNet architecture for predicting cobought events based on product pictures.
%More recently, \citep{zhang2016collaborative} proposes a Deep Network architecture that embeds three types of item information sources, namely Knowledge Base information, text and visual information using a pair-wise ranking loss on the user implicity feedback signal. 
%A second direction of research is on making the representations simple enough to be used for indexing and retrieval, which usually means making them in a smaller space, making them sparse, or both.

The performance of Collaborative Filtering (CF) models is often higher than that of content-based ones but it suffers from the cold-start problem. To take advantage of the best of both worlds, hybrid models use both sources of information in order to make recommendations. One possible way to incorporate product information is using it as side information in the product sequence model, as proposed in Meta-Prod2Vec \citep{vasile2016meta}, leading to better product embeddings for products with low signal (low number of co-occurrences). In this work we continue the investigation of using both types of signal, this time both at training and product recommendation time.

\subsection{Modeling Second Order Interactions}

Modeling second order interactions is a key problem in Machine Learning since the introduction of the kernel trick by \citep{boser1992training}. Recently, real world applications have approximated polynomial kernels by explicit cross features as shown in \citep{chapelle2015simple}. However, this approach can not scale to a very large number of features. Recent work proposed as a solution to factorize the second order terms and introduce Factorization Machines \citep{rendle2012factorization} and Field Aware Factorization Machines \citep{juan2017field} that have achieved state of the art performance in many predictions tasks. Within the deep learning community, \citep{shan2016deep} managed to model second order interactions by merging information through ReLUs. In our paper, we propose the Cross Interaction Unit, a simpler solution that allows fast convergence and good performance with modeling second order interactions. 

%Related work: 
In terms of architecture, our work is also similar to the one proposed by \citep{covington2016deep}, that introduces a scalable solution for video recommendation at YouTube. Unlike their proposed solution, where, in order to support user vector queries, the candidate generation step co-embeds users and items, we are interested to co-embed just the product pairs because for most ecommerce website the number of products is smaller than the number of website users. In our approach, the personalization step can happen after the per-item candidates are retrieved.

\section{Proposed approach: overview}
\label{sec:model}
\subsection{Architecture}
Our proposed approach takes the idea of specializing the input representations to the recommendation task and generalizes it for inputs of different types, in order to leverage all product information and in particular, product images, product title and description text. 

\begin{figure}[t]
\begin{center}
  \includegraphics[scale=0.35]{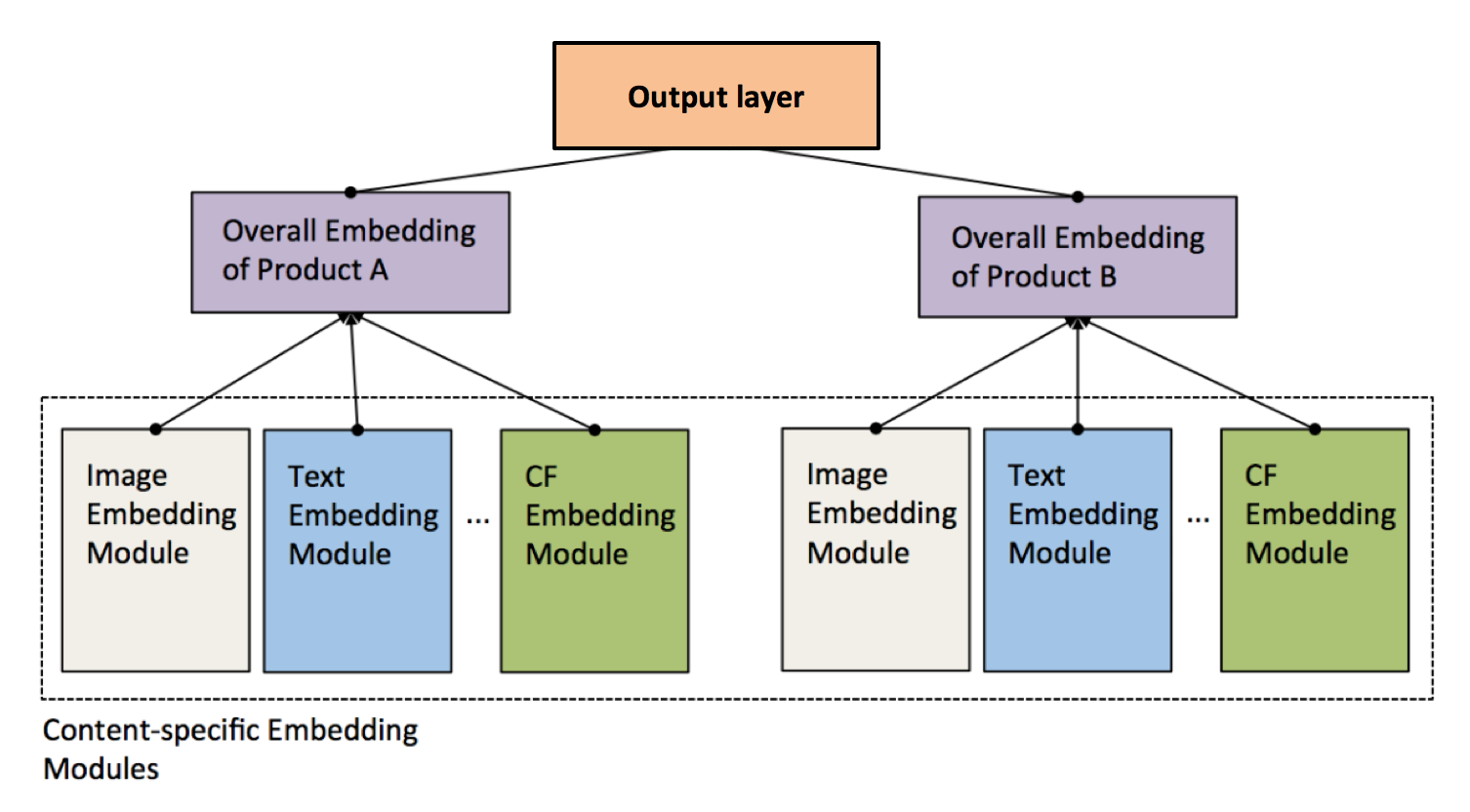}
  \caption{Content2Vec architecture combines content-specific modules to produce embedding vector for each product, then uses these vectors to compute similarities between products. The modality-specific modules are presented in section \ref{sec:Content-specific embedding modules} and the Joint Product Embedding module in Section \ref{sec:architecture:product}}
  \label{fig:content2vec_architecture}
  \end{center}
\end{figure}

The main criteria for the architecture is to allow for the simple plugin of new sources of signal and for the upgrade of existing embedding solutions with new versions (e.g. to replace AlexNet with Inception NN for image processing). As a result, the Content2Vec architecture has three types of modules, as shown in Figure \ref{fig:content2vec_architecture}:
\begin{itemize}[topsep=6pt]
\item \textbf{Content-specific embedding modules} that take raw product information and generate the product vectors. In this paper we cover embedding modules for text, image, categorical attributes and product co-occurrences (description of the differents tested modules in Section \ref{sec:Content-specific embedding modules}). 
\item \textbf{The Joint Product Embedding modules} that merge all the product information into a joint product representation. The two different architectures for this module are detailed in Section \ref{sec:architecture:product}.
\item \textbf{The Output layer} that computes the probability for two products to be cobought or not (this layer is a sigmoid over the inner product between the two unified product embedding vectors)
%\item \textbf{Pair embedding module} that merges the product-to-product interactions and computes the final similarity score. In the case of retrieval-optimized product embeddings, this module becomes the  inner-product between the two items and all interactions between them are to be approximated within the product-level embedding modules.
\end{itemize}

Content2Vec training follows the architecture, learning module-by-module. In the first stage, we initialize the content-specific modules with embeddings from proxy tasks (classification for image, language modeling for text) and re-specialize them to the task of product recommendation. For the specialization task, as mentioned in Section \ref{sec:intro}, we frame the objective as a link prediction task where we try to predict the pairs of products purchased together. We describe the loss function in Section \ref{sec:loss_function} and the different modules in Section \ref{sec:Content-specific embedding modules}.

In the second stage, we concatenate the modality-specific embedding vectors generated in the first stage into a general product vector that is fusioned into a joint representation using the second module. This will be described in depth in Section \ref{sec:architecture:product}.

Finally, in the third stage, given the updated product vectors from stage two, we compute the final probability of being cobought using the output layer. 

\subsection{Learning a pair-wise item distance}
\label{sec:loss_function}
We aim at learning a distance between products that is aligned with the probability of two products being of interest for the same user. The previous work on learning pair-wise item distances concentrated on using ranking loss \citep{mcfee2010metric} or siamese networks with L2 loss \citep{hadsell2006dimensionality}. In \citep{zheng2015logistic},  they introduce the logistic similarity loss :
\begin{align}
L(\theta) = \sum_{ij} -X^{+}_{ij} \log \sigma(sim(a_i,b_j)) -X^{-}_{ij} \log \sigma(-sim(a_i,b_j))
\label{eq:logistic_sim}
\end{align}
where:
\\$\theta = (a_i, b_j)$ is the set of model parameters, where $a_i$ and $b_j$ are the embedding vectors for the products A and B,
\\$X^{+}_{ij}$ is the frequency of the observed item pair $ij$ (e.g. the frequency of the positive pair $ij$),
\\$X^{-}_{ij}$ is the frequency of the unobserved item pair $ij$ (we assume that all unobserved pairs are negatives),
\\ $\sigma$ is the sigmoid function
\\and the similarity distance is defined as:
\begin{align}
sim(a_i, b_j)= \alpha <a_i, b_j> + \beta 
\end{align}
%\\The Negative Sampling loss \citep{mikolov2013distributed} shown in eq. \ref{eq:negsample_sim}  is a fast approximation of the logistic loss. 
%\begin{align}
%L_{NS}(\theta) &= \sum_{ij} -X^{+}_{ij} \log \sigma(sim(a_i,b_j)) \nonumber
%\\&+ \sum_{l=1}^{k} \mathbb{E}_{n_l \sim P_D} \bigg[\log \sigma(-sim(a_i,n_l))\bigg] 
%\label{eq:negsample_sim}
%\end{align}
%where
%\\$P_D$ probability distribution used to sample negative context examples $n_l$,
%\\$k$ is a hyper parameter specifying the number of negative examples per positive example.
%\\
In the following, we detail the different modules used to learn the distance between products. Based on these modules, we can compute some similarities between products based either on their text, their image or their collaborative filtering data. We combine these metrics in the final module. These modules could also be used on their own since they are trained separately to predict whether two products are related or not. 

\section{Content-specific embedding modules}
\label{sec:Content-specific embedding modules}
\begin{figure}[t]
\begin{center}
  \includegraphics[scale=0.4]{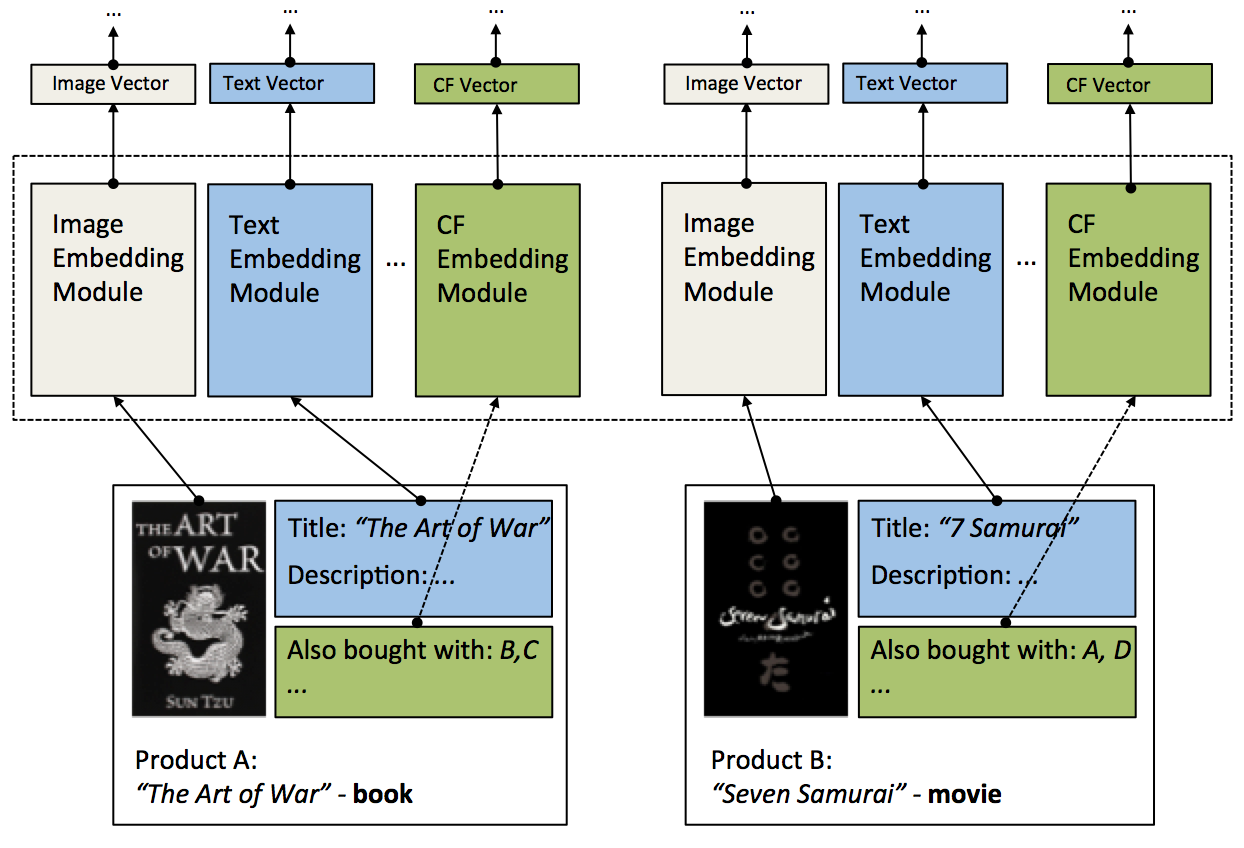}
  \caption{An example of using the content-specific modules to create embedded representations of two products with images, text and CF signal.}
  \label{fig:content2vec_example}
  \end{center}
\end{figure}

Content-specific modules can have various architectures and are meant to be used separately in order to increase modularity. Their role is to map all types of item signal into embedded representations. In Figure \ref{fig:content2vec_example} we give an illustrative example of mapping a pair of products to their vector representations.
In the following we analyze four types of input signal and embedding solutions for each one of them.

\subsection{\textbf{Embedding product images with AlexNet}}

For generating the image embeddings we propose reusing a model trained for image classification, as in previous work by \citep{krizhevsky2012imagenet} and \citep{he2015vbpr}. In \citep{he2015vbpr}, the authors have shown how to use the Inception architecture \citep{szegedy2016rethinking} and specialize it for the product recommendation task. However, the Inception architecture is very deep and requires extensive training time. For ease of experimentation we use AlexNet, which is a simpler architecture that was also a winner on the ImageNet task \citep{krizhevsky2012imagenet} prior to Inception NN or more recently to ResNet \citep{he2016deep}. In section \ref{sec:results} we will show that, even if simpler, when combined with additional product text information, the AlexNet-based solution can perform very well on the recommendation task.

For our experiments, we use the pretrained version of AlexNet available on Toronto's university website.  Our architecture can be seen as a siamese network: the same network is used to build the representation of product A and product B based on their respective image. We specialize the fc7 layer (last layer of AlexNet) to detect product similarities by minimizing the negative sampling loss presented in the previous section. In eq.\ref{eq:logistic_sim}, $a_i$ is the value of the fc7 layer when the input is the image of product A and $b_j$ the same for product B. We only minimize the loss with respect to the weights between fc6 and fc7 since \cite{yosinski2014transferable} proved that it was sufficient to reach good performance (and save a lot of computational resources).  At the end of the optimisation process, we reach visual based embeddings that are specialized to predict if two products could be co-bought or not.
 
\subsection{\textbf{Embedding product text with Word2Vec and CNN on sentences}}

With this module, we want to compute products similarities based on their text descriptions. The module is trained in two steps. First, we train a Word2Vec \cite{mikolov2013distributed} model in order to reach a representation of each word of the catalogue. Then, we train a TextCNN \citep{kim2014convolutional} that combines the words representation to build a product text embedding trained to detect whether two products were cobought or not. To the best of our knowledge, this is the first attempt to employ a TextCNN architecture for the task of product recommendation.

For generating word embeddings, we propose reusing Word2Vec, a model for generating language models that has been employed in a various of text understanding tasks. More recently, it has been shown in \citep{glove} that Word2Vec is closely linked with matrix factorization techniques applied on the word co-occurrence matrix. We chose to train Word2Vec on the entire product catalog text information and not use an available set of word embeddings such as the one created on the Google Corpus. The main reason is that the text distribution within product descriptions is quite different from the general distribution. For example the word \emph{'jersey'} has a very different conditional distribution within the product description corpus versus general online text.

TextCNN offers a simple solution for sentence-level embeddings using convolutions. The convolutions act as a form of n-gram filters, allowing the network to embed sentence-level information. We learn the filters by minimizing the negative sampling loss to predict if two products were cobought. The trainable parameters of the model are the weights in the convolutional layer that we call filters. 

We keep fixed the word embeddings that were trained through Word2Vec as recommended in \citep{kim2014convolutional}. Using embeddings helps to generalize to words that were not frequent in the training set. The architecture has the advantage to allow to choose an arbitrary length for the text input. The two siamese networks are fed by the word embeddings of the first 10 token of the concatenated product title and description of product A and B.  We only tested with keeping the first 10 tokens for ease of experiments (it is straightforward to extend to a higher number of considered tokens). We provide a comparison of performance with the image-based module in the experiments section. 

\subsection{\textbf{Embedding product co-occurrences with Prod2Vec}}
Prod2Vec \citep{prod2vec} is an extension of the Word2Vec algorithm to product shopping sequences. As a result, Prod2Vec can be seen as a matrix factorization technique on the product co-occurence matrix (see \cite{glove}). 
In Content2Vec, the Prod2Vec-based similarity contains all of the information that can be derived from the sequential aspect of the user behavior, without taking into account the per-product meta-data.
%\todo{TODO: Mention no pretrain!}

\subsection{\textbf{Embedding categorical product meta-data with Meta-Prod2Vec}} 
Meta-Prod2Vec \citep{vasile2016meta} improves upon Prod2Vec by using the product meta-data side information to regularize the final product embeddings. In Content2Vec, we can use the similar technique of co-embedding product categorical information with product ids to generate the embedding values for the categorical features. 
%\todo{TODO: Mention no pretrain!}

\section{The Joint Product Embedding Module}
\label{sec:architecture:product}

With these different modules, we can build several product embeddings that are modality-specific and specialized for the task of product recommendation. In the next stage, we combine them in order to build a unified representation of the product. This representation will consider all the signals available on the product: text, image and collaborative filtering data.

\subsection{\textbf{Joint Product Embedding with performance constraints}}

As stated in Section \ref{sec:intro}, the function of the product embedding module is two-fold: first, to model all interactions that exist between the modality-specific embeddings with respect to the final optimization objective, and second, to approximate interaction terms between the products that cannot be explained by a linear combination of the modality-specific similarities. With this in mind, we introduce a new type of learning unit, the \emph{Cross Interaction Unit} (eq. \ref{eq:pair_res_unit}).
%, which similarly to the original \emph{residual unit} introduced in \cite{he2016deep} and allow the layers to learn incremental, i.e. residual representations.
%Add something on FFMs and polynomial kernels
%In \cite{hardt2016identity} the authors motivate the use of residual units as helping preserve the representations learned in the previous layers. In our case we are interested in preserving the specialized image and text representations and learn an additional representation for their interactions.  We decided to use the term \emph{residual} in describing our new pairwise embedding layer based on the similarity in motivation with the original residual unit that was introduced to help the system approximate the identity and allows the new layers to focus on the remaining unexplained error. Of course, in our case, the layer does not serve the same practical purpose as in the original ResNets architectures that use the unit mostly for training very deep networks.
%\begin{figure}[h!]
%\begin{center}
%  \includegraphics[scale=0.3]{pairwise_res_unit.png}
%  \caption{Pairwise Residual Unit}
%  \label{fig:pairwise_res_unit}
%  \end{center}
%\end{figure}
%In the original use case, a residual layer can be defined as: 
%\begin{eqnarray*}
%y = F(X) + X
%\label{eq:res_unit}
%\end{eqnarray*}

We note $X_1$ (respectively $X_2$) the two product embedding vectors (obtained by stacking the modality-specific vectors):
\begin{align}
X_1 = [X_1^{mod_1}, X_1^{mod_2}, ..., X_1^{mod_N}]
\end{align}
where ${mod_i}$ are the different modalities considered by the architecture. We define the \emph{Cross Interaction Unit} as:
\begin{align}
y = sim(F(X_1),F(X_2)) + sim(X_1,X_2)
\label{eq:pair_res_unit}
\end{align}
where:
\\$sim(.,.)$ is a similarity function over two embedding vectors $X_1$, $X_2$,
\\$F(x)$ is a ReLu layer.

The F function can be seen as a module that will use information coming from all the modalities to explain similarities that could not be explained by simply using their linear contribution. 
%Unlike the fully connected compressed layer, it does not need to focus on explaining similarities that are already explained by one modality. 

To be able to measure the incremental value of introducing this residual vector we introduce a baseline architecture that computes the final prediction based on the linear combination of the modality-specific similarities denoted by \emph{Content2Vec-linear} with the associated similarity function defined in eq. \ref{eq:sim_c2v}.

\begin{eqnarray}
sim_{c2v-lin}(X_1, X_2) = \sum_{m \in Modalities} w_m sim_{m}(X_1^m, X_2^m)
\label{eq:sim_c2v}
\end{eqnarray}

Under this notation, the CIU-based architecture denoted as \emph{Content2Vec perf} minimizes the logistic loss with the similarity function defined in eq. \ref{eq:sim_c2v_res}.

\begin{eqnarray}
sim_{c2v-res}(X_1, X_2) = \sum_{m \in (Modalities + Residual)} w_m sim_{m}(X_1^m, X_2^m)
\label{eq:sim_c2v_res}
\end{eqnarray}

In order to learn the residual vector, we keep fixed the modality-specific similarities and co-train the final weights of each of the modalities together with the product-specific residual layer. For example, in the case of using only image and text signals, our final predictor can be defined as in the following equation where $P_{txt}$ and $P_{img}$ are pre-set and $w_{txt}$, $w_{img}$, $w_{res}$ and $P_{res}$ are learned together:

\begin{align}
P(pos|X_1, x_2) &= \sigma\bigg(w_{txt} P_{txt}(pos|X_1^{txt},X_2^{txt}) 
\\&\quad + w_{img}P_{img}(pos|X_1^{img},X_2^{img}) \nonumber
\\&\quad+ w_{res}P_{res}(pos| X_1^{res}, X_2^{res})\bigg) \nonumber
\label{eq:example_res}
\end{align}

with:
\begin{equation}
P_{res}(pos| X_1, X_2) = \alpha <F([X_1^{txt},X_1^{img}]),F([X_2^{txt},X_2^{img}])> + \beta
\end{equation}
In Section \ref{sec:results} we compare the performance of \emph{Content2Vec-perf} and \emph{Content2Vec-linear} and show that, as expected, the proposed architecture surpasses the performance of the linear model, while allowing for a retrieval-based candidate scoring solution.

\subsection{\textbf{Joint Product Embedding with size constraints}} 

One of the main objectives of the paper is to investigate architectures that can reach a unified product embedded representation that is optimized for the task of retrieval-based product recommendation. In order to enable an efficient retrieval task (\cite{hubara2016quantized} and \cite{babenko2016pairwise}), we implement an architecture that places a constraint on the size of the final vector representing the product of small dimension (we set arbitrarily the allowed dimension to 200 in our case). To this end, we represent the product representation module by a fully connected ReLU layer that compresses the representation of the product into a vector of size 200. This layer takes as input a concatenated version of all the vectors available on the products from the modality-specific modules. After this compression layer, we reach one unified vector for product A and one unified vector for product B. These vectors are then used to compute product similarities by computing the inner product between two products representations. In the following, we will refer to this architecture by \emph{Content2Vec-compressed}. 

The product representations coming from the modality-specific modules are already relevant w.t.r.t the final task of product pair prediction, as shown by the performance of Image and TextCNN baselines in section \ref{sec:results}. In order to avoid losing information coming from the concatenated vector representation, different initializations of the fully connected layer are possible. Of course, the initialisation that would keep the incoming information from the trained modalities networks would be a layer close to the identity matrix . One way to approximate this objective would be to compute the PCA over vector representation of the full training dataset, but this is computationally expensive. We propose a simpler initialization that considers 200 random products (the number of products considered is defined by the dimension we want to impose to the final vector) from the training dataset and uses their representation as rows of the matrix. Hence, at the beginning, the layer will compute the similarities between each product and these 200 source products.

\begin{table*}
\centering
\begin{tabular}{lrrr} 
\toprule
\textbf{Model trained on Books dataset} & \textbf{Books} & \textbf{Movies} & \textbf{Mixed}\\
\toprule
\textit{Modality Baselines}\\
ImageCNN     & 81\%       & 78\%        & 64\%\\
TextCNN                  & 72\%       & 79\%        & 76\%\\
\midrule
\textit{Fusion Baselines}\\
Fusion-linear        & 83\%       & \textbf{83\%}        & 76\%  \\
Fusion-crossfeat     & 86\%       & \textbf{83\%}        & \textbf{83\%}  \\
\midrule
\textit{Our approaches}\\
Content2Vec-compressed        & 85\%       & 81\%        & 64\%       \\
Content2Vec-perf           & \textbf{89\%}       & \textbf{83\%}        & 77\% \\
\bottomrule
\end{tabular}

\begin{tabular}{lrrr} 
\toprule
\textbf{Model trained on Movies dataset} & \textbf{Movies} & \textbf{Books} & \textbf{Mixed}\\
\toprule
\textit{Modality Baselines}\\
ImageCNN  & 92\%        & 59\%               & 60\%       \\ 
TextCNN       & 90\%        & 63\%               & 65\%       \\ 
\midrule
\textit{Fusion Baselines}\\
Fusion-linear       & 94\%   & \textbf{64\%}               & 65\%       \\ 
Fusion-crossfeat   & 94\%   & 62\%               & 63\%       \\
\midrule
\textit{Our approaches}\\
Content2Vec-compressed       & \textbf{95\%}      & 54\%             & 58\%       \\
Content2Vec-perf          & \textbf{95\%}     & 60\%              & \textbf{66\%}       \\
\bottomrule
\end{tabular}
\centering
\caption{Comparative results in terms of AUC between the different architectures on the hard cold start dataset (test set: same category as during training, different category as during training, mixed category)}
\label{table:hard_cold_start_results}
\label{table:hard_cold_start_results}
\end{table*}

\section{Experimental Results}
\label{sec:experimentsetup}

\subsection{Dataset}

We perform our evaluation on the publicly available Amazon dataset \citep{mcauley2015image} that represents a collection of products that were co-bought on the Amazon website. Each item has a rich description containing product image, text and category.
In terms of dimensionality, the dataset contains around 10M pairs of products.  
We concentrate on the subgraph of Book and Movie product pairs, because both categories are large and they have a reasonable sized intersection. This allows us to look at recommendation performance on cross-category pairs (to evaluate a model trained only on Book pairs on predicting Movie co-bought items) and mixed category pairs (to evaluate the models on Book-Movie product pairs).
\\
\\
Based on the full Book \& Movies data we generate two datasets with different characteristics:
\begin{itemize}
\item The first dataset simulates a \textbf{hard cold start regime}, where all product pairs used in validation and testing are over products unseen in training. This tests the hardest recommendation setup, where all testing data is new. We decided to bench all of our hyperparameters on this regime and use the best setup on all datasets, since tuning on the harder dataset ensures the best generalization error (results shown in Table \ref{table:hard_cold_start_results}).  
\item The second dataset simulates a \textbf{soft cold start regime}, where some of the products in the test set are available at training time. The dataset is generated by taking the top 200k most connected products in the original dataset and sampling 10\% of the links between them (results shown in Table \ref{table:soft_cold_start_results}).  %This dataset tests the increase in performance of a hybrid model that combines the CF and Content signal in a setup that is close to what we observe in real life applications 
\end{itemize}
\paragraph{Hyper-parameters}
We fixed the sizes of embedding vectors for image CNN module to 4096 hidden units, for text CNN module to 256, for Prod2Vec module to 50, for residual representation to 128. For optimization, we use the Adam algorithm (\citep{kingma2014adam}) and we manually set the initial learning rate based on the validation set performance. The batch sizes vary for different datasets. We train all the models until validation set performance stops increasing.

\paragraph{Loss} 
Instead of minimizing the logistic loss, we minimize the Negative Sampling loss \citep{nce} which is a fast approximation of the logistic loss. The prediction step can scale up to a large number of items, by using all positive pairs and sampling the negatives on the fly.

\paragraph{Evaluation task}
We evaluate the recommendation methods on the product link prediction task, similar to \citep{he2015vbpr}. We consider the observed product pairs as positive examples and all unknown pairs as negatives. We generate negative pairs according to the frequency of the products in the positive pairs (negative examples between popular products are more likely to be generated) with a positive to negative ratio of 1:2.

\paragraph{Evaluation metrics}
For the link prediction task, we use the Area Under Curve of the Precision/Recall ( curve as our evaluation metric.
\begin{table*}
\centering
\begin{tabular}{lr} 
\toprule
\textbf{Recommendation Models} & \textbf{Test} \\
\toprule
\textit{Baselines}\\
ImageCNN                & 80\% \\ 
TextCNN               & 78\% \\ 
Prod2vec                & 86\% \\ 
Fusion-linear        & 88\% \\ 
Fusion-linear+       & 89\% \\ 
\midrule
\textit{Without Prod2vec signal}\\
Content2vec-compressed  & 87\% \\ 
Content2vec-perf           & 89\% \\ 
\midrule
\textit{With Prod2vec signal}\\
Content2vec-compressed+ & 89\% \\ 
Content2vec-perf+         & \textbf{92\%} \\
\bottomrule
\end{tabular}
\centering
\caption{Comparative results in terms of Area Under Precision-Recall Curve (AUPRC) between the different architectures on the soft cold start dataset}
\label{table:soft_cold_start_results}
\end{table*}
\paragraph{Baselines}
We introduce several baselines and compare their performances with our proposed architectures: 
\begin{itemize}
%\item \textit{ImageCNN W-cosine}: the method of predicting co-bought products based on weighted innerproduct of the product image embeddings.
\item \textit{ImageCNN}: prediction based on specialized image embeddings similarity (that previously showed state-of-the-art results on the Amazon dataset \cite{mcauley2015image})
\item \textit{TextCNN}: prediction based on specialized text embeddings similarity
\item \textit{Prod2Vec}: prediction based on the product vectors coming from the decomposition of the co-purchase matrix
\item \textit{Fusion-Linear}: prediction based on the linear combination of text and image similarities
\item \textit{Fusion-Crossfeat}: prediction based on the linear combination of discretized image and text similarities and their conjuctions:  we bucketize the text and image-specific similarity scores and create explicit feature conjunctions between them. 
%\item \textit{Content2Vec-embedpairs}: prediction based on the linear combination of text and image similarities and a pair-level residual component
\end{itemize}	

\paragraph{Our approaches} 
\begin{itemize}
\item \textit{Content2Vec-compressed}: prediction based on the compressed version of the product representation
\item \textit{Content2Vec-perf}: prediction based on the linear combination of text and image similarities plus product-level residual vectors similarities
\item \textit{Content2Vec+}: prediction based on the ensemble of Prod2Vec and Content2Vec models
\end{itemize}

\subsection{Results}
\label{sec:results}

The two following tables (Tables \ref{table:hard_cold_start_results}, \ref{table:soft_cold_start_results}) correspond to the two types of dataset we consider: the hard cold start dataset where all product pairs used in validation and testing are over products unseen in training and the soft one where some of the products in the test set are available at training time.

\subsubsection{\textbf{Difference of performance between baselines}} 
To the best of our knowledge, no study has been made so far on the performance of TextCNN in the recommendation system setting. We observe that it is slightly worse than ImageCNN when evaluated on the same product category it was trained. On the mixed dataset (pairs of products from both Books and Movies), TextCNN generalize better than ImageCNN. We cannot bench Prod2Vec on the hard cold start dataset since no collaborative filtering data on the test products were available at training time. When good collaborative filtering data is available, Prod2Vec outperfoms both TextCNN and ImageCNN but the performance of Prod2Vec strongly depends on the degree of connectivity of the products graph. For new products, Prod2Vec can not be used. 

\subsubsection{\textbf{Combining product signals}}

The results show that first our two main proposed method \emph{Content2Vec-compressed} and \emph{Content2Vec perf} can leverage the additional signal provided by each of the input modalities in a joint manner and leads to significant gains in AUC versus the one-signal baselines (ImageCNN, TextCNN) and their linear combination (\emph{Content2Vec-linear}). 

From the point of view of robustness, \emph{Content2Vec-perf} learns product representations that perform better than the baseline methods on out-of-sample recommendations such as cross-category pairs and mixed-category pairs (Table \ref{table:hard_cold_start_results}). However, there still exists a gap for improvement. 

\subsubsection{\textbf{Tradeoff between performance and reaching a compressed product representation}}

 \emph{Content2Vec-compressed} performs slightly worse than  \emph{Content2Vec-perf} but it uses a more compressed representation than enables a more efficient retrieval system. We also remark that the distance and the representation learned by  \emph{Content2Vec-compressed}  is more category-specific that \emph{Content2Vec-perf} since \emph{Content2Vec-perf} is performing better in the cross-category setting. Besides, the training time before reaching a good  \emph{Content2Vec-compressed} model is higher than \emph{Content2Vec-perf}. Hence, there exists a clear trade-off in order to choose between \emph{Content2Vec-perf} and \emph{Content2Vec-compressed}: in terms of prediction performance, training time and cross-category efficiency,  \emph{Content2Vec-perf} has better results, while \emph{Content2Vec-compressed} still achieves good prediction performance and offers a compressed product representation.
 
 \subsubsection{\textbf{Incorporating Prod2Vec signal}}

\emph{Content2Vec-perf+}, our proposed hybrid architecture that combines content and CF signal achieves better performance than the content and CF-only models. It confirms that even on a setting where good collaborative data are available, the representation can be improved by using all other signals. We find also interesting that our content-based method \emph{Content2Vec-perf} has a similar performance than a collaborative-filtering based method such as Prod2Vec.  

%\subsubsection{\textbf{Difference of performance between products and pair embeddings}}

%We observe that adding an additional layer that represents pair-level interactions does not lead to big improvements in either of the two models we investigated (Content2Vec-crossfeat,embedpairs), confirming that a product retrieval-based recommender system can achieve state-of-the-art results.

\section{Conclusions}
\label{sec:conclusions}
In this paper, we propose \emph{Content2vec}, a new product representation architecture which addresses most of the requirements outlined in the \emph{Joint Product Representation} task. It generates relevant representations by optimizing for the target offline metric i.e AUC of hold-out product pairs prediction and covers all input signal by using a representation module for each type of signal. It also offers cross-modality expressiveness by the introduction of the product embeddings modules and optionally pair-wise expressiveness in the pair embedding module, passes robustness checks by performing better than baselines on hard cold start and cross-category evaluation tasks, and offers the possibility for retrieval-optimized vectors with the \emph{Content2vec-compressed} version.
\\
\\
This work has several key contributions: We develop a method that is able to use all product signal for the task of product recommendation using a modular architecture that can leverage fast evolving solutions for each type of input modality. We define a set of requirements for evaluating the resulting product embeddings and show that our method leads to significant improvements over the single signal approaches on hard recommendation situations such as cold-start and cross-category evaluation. We show how to build a compressed product representation that is able to take into consideration all signal available on the product to perform well on some hard cold start setting and improve the collaborative filtering representation in a normal recommendation scenario. Finally, in order to model the joint aspects of the product embedding with keeping some linearities in the model we introduce a new type of learning unit, named \emph{Cross Interaction Unit} and show the resulting gains on a real product co-purchases dataset.
\\
\\
For the next steps, we would like to improve \emph{Content2vec} for cross-category tasks and impose some sparsity constraints on product representations to increase the performance of the final product retrieval system. 
\newpage

\bibliographystyle{ACM-Reference-Format}
\bibliography{literature} 
\newpage
\appendix 
\section{Infrastructure of the recommendation system} 
\begin{figure}[h!]
\begin{center}
  \includegraphics[scale=0.43]{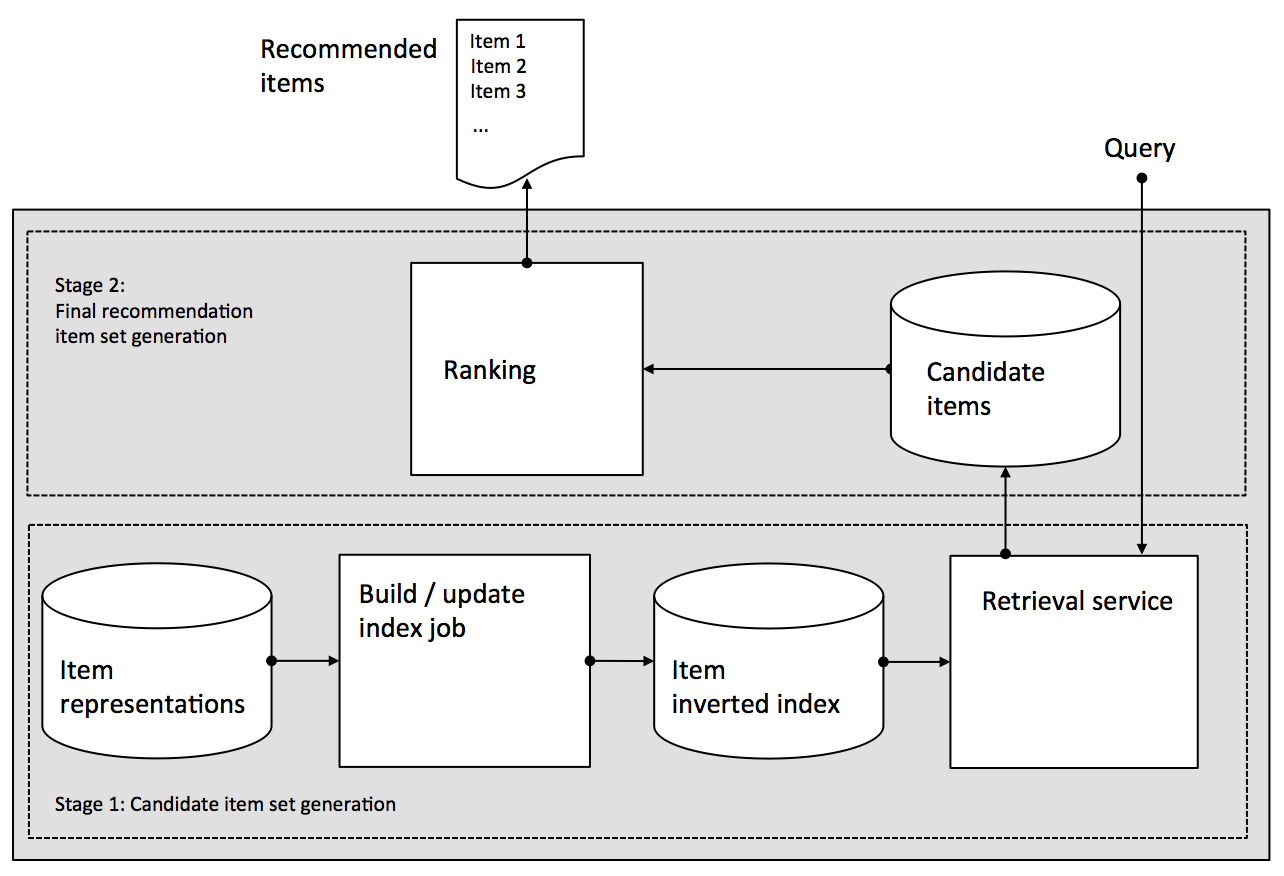}
  \caption{2-Stage Recommender System Architecture.}
  \label{fig:criteo_recsys}
  \end{center}
\end{figure}

\end{document}